\documentclass[journal]{IEEEtran}
\usepackage{xurl}      
\usepackage{cite}
\usepackage{amsmath,amssymb,amsfonts}
\usepackage{algorithm}
\usepackage{algorithmic}
\usepackage{graphicx}
\usepackage[mathlines,switch]{lineno}
\usepackage{multicol,multirow}
\usepackage{stfloats}
\usepackage{booktabs}
\usepackage{tabularx}
\usepackage{array}
\newcolumntype{Y}{>{\raggedright\arraybackslash}X}

\title{Three Stages of ISAC Signal Design: Principles, Methods and Standardization}

\author{Zhiyong Feng,~\IEEEmembership{Senior Member, IEEE},
Zhiqing Wei,~\IEEEmembership{Member, IEEE},
Haotian Liu,~\IEEEmembership{Graduate Student Member, IEEE},
Yichen Wu,~\IEEEmembership{Graduate Student Member, IEEE}, 
Lin Wang,~\IEEEmembership{Graduate Student Member, IEEE}, 
Yuhan Long

\thanks{Zhiyong Feng, Zhiqing Wei, Haotian Liu, Yichen Wu, Lin Wang, and Yuhan Long are with the Key Laboratory of Universal Wireless Communication, Ministry of Education, Beijing University of Posts and Telecommunications, Beijing 100876, China (emails: \{fengzy; weizhiqing; haotian\_liu; wuyichen; wlwl; 202411020\}@bupt.edu.cn). \textit{Corresponding author}: Zhiqing Wei, Haotian Liu.}}

\begin{document}

\maketitle

\begin{abstract}
Emerging scenarios, including low-altitude economy and intelligent transportation,
require ultra-reliable communication and high-precision sensing,
making integrated sensing and communication (ISAC) a key enabler for sixth-generation (6G) networks.
The ISAC signal serves as the carrier for both communication and sensing,
and its design directly affects the overall performance of ISAC system.
This article presents a progressive roadmap of ISAC signal design.
First, we establish a multi-dimensional performance metric framework of ISAC signal covering communication, sensing, and radio-frequency (RF) performance.
Second, we introduce a three-stage ISAC signal design paradigm,
namely \textsc{Unleash Potential}, \textsc{Expand Dimensions}, and \textsc{Deepen Collaboration}.
This paradigm progresses from the exploitation of existing signals, 
through the design of new waveform bases and multi-dimensional signal structures,
to the ISAC signal design with multi-node cooperation.
Finally, we propose a corresponding standardization roadmap
that evolves from compatibility with existing communication systems toward native and cooperation-enabled ISAC.
The resulting framework provides practical guidance for ISAC signal design and the standard evolution toward 6G.
\end{abstract}

\begin{IEEEkeywords}
Integrated sensing and communication; Signal design; Standardization.
\end{IEEEkeywords}

\section{Introduction}\label{sec:I}

The rapid evolution of sixth-generation (6G) networks is creating new demands for wireless connectivity and environmental awareness.
Emerging applications, including low-altitude economy and intelligent transportation,
require wireless networks to support high-rate and low-latency communications together
with high-precision, wide-area, and flexible sensing of the physical environment~\cite{wei2024deep}.
As a key enabler of 6G,
integrated sensing and communication (ISAC) supports information transmission and environment sensing by sharing spectrum, hardware,
and signal processing resources, thus improving overall resource efficiency and enabling mutual enhancement between communication and sensing~\cite{liu2022integrated}.
As the carrier for both communication and sensing,
ISAC signal has attracted considerable attention and has become a key factor in improving overall performance.
Due to hardware constraints, especially the limited power budgets and power-amplifier capabilities of the base station (BS) and user equipment (UE), 
ISAC signal design should jointly balance communication, sensing and radio-frequency (RF) performance. 
Consequently, several key challenges arise in ISAC signal design.

\begin{itemize}
  \item \textbf{Lack of multi-dimensional Performance Metrics:}
The existing ISAC signal design has considered the tradeoffs between communication and sensing performance.
However, multi-dimensional performance metrics, including communication, sensing, and RF performance, have not yet been jointly characterized,
limiting the effectiveness of ISAC signal design.
  \item \textbf{Limited Compatibility to Mobile Communication Systems:}
Conventional mobile communication systems are designed primarily for communication services.
As a result, their RF front ends and baseband processing architectures cannot be readily adapted to ISAC signal processing.
A practical and evolutionary roadmap for ISAC signal design is required to enable smooth compatibility with existing and future mobile communication systems.
  \item \textbf{Difficulty in Reaching the Consensus in Standardization:}
Existing studies have developed various ISAC signal design approaches tailored to different scenarios and requirements.
Reaching consensus on a unified standardization framework remains a challenge.
\end{itemize}

Existing studies related to these challenges are summarized as follows.

\textbf{ISAC Performance Metrics:} 
Wei et al.~\cite{wei2023waveform} used mutual information to characterize the communication and sensing performance of ISAC systems.
Liu et al.~\cite{liu2022integrated} characterized the communication and sensing performance boundary using the Cramér–Rao bound (CRB)–rate region.
However, RF performance, such as the peak-to-average power ratio (PAPR),
has not been fully incorporated into existing performance evaluation frameworks.

\textbf{ISAC Signal Design:} 
The communication-centric ISAC signal design schemes are mainly built on standard communication waveforms such as
orthogonal frequency-division multiplexing (OFDM) and discrete Fourier transform-spread orthogonal frequency-division multiplexing (DFT-s-OFDM). 
In addition, new waveforms such as orthogonal time frequency space (OTFS) and affine frequency-division multiplexing (AFDM) have been investigated with the aim of improving adaptability to time-varying channels~\cite{rou2024orthogonal}.
However, existing ISAC signal design studies still lack a systematic roadmap that supports long-term technological evolution.

\textbf{ISAC Signal Standardization:} 
According to the ongoing progress of the 3rd Generation Partnership Project (3GPP) Release 20 and Release 21,
ISAC signal standardization is still at an early stage,
where sensing is mainly enabled by reusing existing 5G New Radio (NR) waveforms and reference signals~\cite{eren2025integrated}.
Nevertheless, a well-defined standardization roadmap towards multi-node cooperation, flexible communication-sensing resource sharing,
and native ISAC signal design has yet to be established.

To this end, this article develops a three-stage progressive framework for ISAC signal design that covers three complementary dimensions:
performance evaluation, signal design, and standardization.
The main contributions are summarized as follows.
\begin{itemize}
    \item \textbf{Performance Tradeoffs:}
    We establish a unified framework to characterize the tradeoffs among communication, sensing, and RF performance, which provides guidelines for ISAC signal optimization.
    \item \textbf{Three-Stage Signal Design:}
    We propose a progressive design roadmap comprising \textsc{Unleash Potential}, \textsc{Expand Dimensions}, and \textsc{Deepen Collaboration}.
    These stages, respectively, leverage existing waveforms, develop new waveform bases, and extend ISAC signal design from single-node operation to multi-node cooperation.
    \item \textbf{Standardization Roadmap:} 
    We outline a practical three-stage evolution path toward 6G, ranging from compatibility with existing 3GPP systems to native and cooperative ISAC.
\end{itemize}

This article is organized as follows.
Section II introduces the fundamental concepts of ISAC systems.
Section III investigates the multi-dimensional performance tradeoffs in ISAC signal design.
Section IV presents the proposed progressive roadmap for ISAC signal design.
Section V elaborates on the evolution and progress of standardization of ISAC signal design.
Finally, Section VI concludes this article.

\section{Basic Concepts of ISAC Systems}

This section introduces the sensing modes of ISAC,
the performance metrics for ISAC signal design,
and the signal design requirements for typical application scenarios.

\subsection{Sensing Modes}
ISAC sensing modes can be broadly classified into single-node sensing and cooperative multi-node sensing.
Different sensing modes impose different requirements and challenges for ISAC signal design.

\subsubsection{Single-Node Sensing}
Single-node sensing can be classified into downlink sensing and uplink sensing.
In downlink sensing, a BS transmits the ISAC signal and processes the resulting echoes.
Compared with UE, the BS generally features a larger power budget and more capable RF hardware.
The primary design challenge therefore lies in reconciling the random data‑bearing structure favored by communications with the narrow main‑lobe and low sidelobe characteristics demanded for accurate sensing.

In uplink sensing, an UE transmits ISAC signals, while a BS receives and processes the reflected signals.
The limited transmit power and power-amplifier capability of the UE directly constrain both sensing coverage and accuracy.
Consequently, the core challenge is to balance the sensing performance, communication reliability, and RF efficiency subject to the practical hardware limitations of the UE.

\subsubsection{Cooperative multi-node sensing}
Compared to single-node sensing, cooperative sensing expands the sensing coverage and improves the estimation accuracy~\cite{wei2024deep}.
However, such performance gains require coordinated multi-node signal design.
In multi‑BS cooperative sensing, multiple BSs jointly transmit ISAC signals, collect echo signals, and perform sensing‑information fusion.
A major difficulty arises from the separation of echoes originating from different transmitters, propagation paths, and targets in the presence of inter-BS interference.
Consequently, the core problem is the design of mutually orthogonal or separable ISAC signals under resource constraints, 
to enable reliable echo separation for multi‑BS cooperative sensing.

In BS‑UE cooperative sensing, multiple UEs transmit uplink ISAC signals, and BS processes the echo signals.
Consequently, it is a critical issue to implement efficient joint resource allocation and optimize overall communication and sensing performance under limited UE transmit capabilities.

\subsection{Performance Metrics for ISAC Signals}
The performance of ISAC waveforms should be jointly evaluated in the sensing, communication, and RF domains.
Table~\ref{tab:isac_metrics} summarizes the corresponding performance metrics.

\subsubsection{Sensing Performance Metrics}
Sensing performance metrics characterize the ability of the ISAC waveform to detect targets and acquire information about the surrounding environment.
For signal design, sensing performance is primarily evaluated in terms of target detectability and resolution, correlation and sidelobe characteristics, as well as parameter‑estimation accuracy.
Representative performance metrics of sensing include sensing resolution, peak sidelobe level ratio (PSLR), Cramér–Rao lower bound (CRLB), and mean‑squared error (MSE), 
as listed in Table~\ref{tab:isac_metrics}.

\subsubsection{Communication Performance Metrics}

Communication performance metrics quantify the information‑transmission efficiency and reliability of an ISAC signal under resource competition, waveform distortion, and time‑varying channel conditions.
Representative performance metrics include spectral efficiency (SE), bit error rate (BER), error vector magnitude (EVM), and robustness against multipath and Doppler effects, as summarized in Table~\ref{tab:isac_metrics}.

\subsubsection{RF Performance Metrics}

RF performance metrics characterize the compatibility of an ISAC signal with practical transmission hardware.
They evaluate the performance of signal from the perspective of power‑amplifier efficiency, nonlinear distortion, and out‑of‑band emissions.
Representative metrics include peak‑to‑average power ratio (PAPR), cubic metric (CM), and adjacent channel leakage ratio (ACLR).
These performance metrics are of particular importance for power‑constrained uplink transmission and UE‑assisted sensing scenarios.

\begin{table*}[!t]
\centering
\caption{Core Metrics for ISAC Signal Design}
\label{tab:isac_metrics}
\small
\renewcommand{\arraystretch}{1.12}
\begin{tabularx}{0.8\textwidth}
{@{}p{0.13\textwidth} p{0.22\textwidth} X@{}}
\toprule
\textbf{Dimension} & \textbf{Metric} & \textbf{Design Relevance} \\
\midrule

\multirow{4}{*}{\textbf{Sensing}}
& Resolution
& Target separability and high-resolution detection or imaging~\cite{wei2024multiple}. \\

& PSLR
& Sidelobe suppression and mitigation of weak-target masking~\cite{Long2026}. \\

& CRLB
& Theoretical limit of parameter-estimation accuracy~\cite{liu2022integrated}. \\

& MSE
& Practical parameter-estimation accuracy~\cite{wei2023waveform}. \\

\midrule

\multirow{4}{*}{\textbf{Communication}}
& SE
& Bandwidth utilization and communication efficiency. \\

& BER
& Transmission reliability~\cite{eren2025integrated}. \\

& EVM
& Modulation fidelity and link distortion~\cite{Long2026}. \\

& Multipath/Doppler robustness
& Adaptability to mobility and time-varying channels~\cite{koivunen2024multicarrier}. \\

\midrule

\multirow{3}{*}{\textbf{RF}}
& PAPR
& Envelope fluctuation and power-amplifier backoff~\cite{Long2026}. \\

& CM
& Sensitivity to power-amplifier nonlinearity~\cite{Long2026}. \\

& ACLR
& Out-of-band emissions and adjacent-channel interference. \\

\bottomrule
\end{tabularx}
\end{table*}

\subsection{ISAC Scenarios and Requirements}
This subsection discusses the requirements for communication, sensing, and RF performance in several representative ISAC scenarios.

\subsubsection{Intelligent Transportation}
In intelligent transportation systems, ISAC supports reliable communication and high‑resolution environmental sensing.
Consequently, a low BER is required for reliable information transmission, while high resolution and low MSE are needed for accurate environmental sensing.
Hence, careful tradeoffs among communication spectral efficiency, transmission reliability, and sensing accuracy are required in ISAC signal design.

\subsubsection{Low-Altitude Economy}

In low‑altitude economy scenarios, ISAC enables reliable command and sensing data transmission for unmanned aerial vehicle (UAV) swarms.
However, constraints on battery capacity, payload capacity, and platform size make energy-efficient RF operation essential, requiring signal designs that maintain communication reliability while reducing PAPR and RF front-end power consumption.

\section{Performance Tradeoffs of ISAC Signals}

\begin{figure*}[htbp]
\centering
\includegraphics[width=0.82\textwidth]{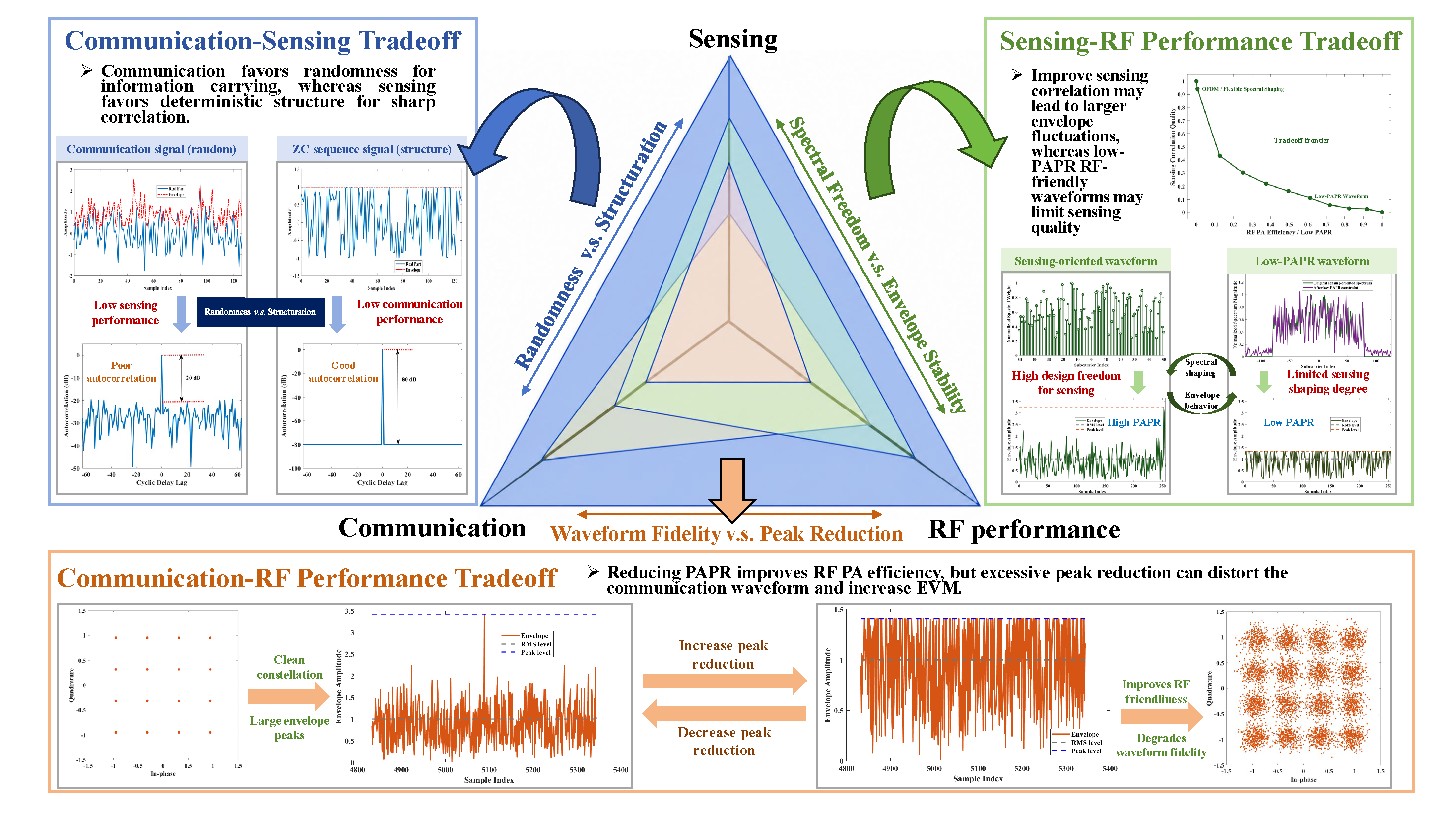}
\caption{Performance tradeoffs in ISAC signal design.}
\label{fig:fig1}
\end{figure*}

As illustrated in Fig.~\ref{fig:fig1}, ISAC signal design involves multi-dimensional tradeoffs among communication, sensing, and RF performance.  
This section clarifies the interactions among these three performance metrics, thereby providing a theoretical foundation for ISAC signal design.


\subsection{Tradeoff between Sensing and Communication Performance}
The fundamental tradeoff between communication and sensing arises from the intrinsic conflict in signal structures.
Communication needs to accommodate random information-bearing symbols, while sensing benefits from deterministic or highly structured signals with favorable correlation characteristics and low sidelobe levels.

One viable approach is to directly optimize the sensing‑related properties of communication signals.
Representative techniques include embedding phase factors into single‑carrier signals or introducing controlled symbol deviations to suppress ranging sidelobes.
However, these techniques may incur extra signaling overhead or increase BER.
Alternatively, time‑frequency resources can be jointly allocated between communication and sensing.
Allocating more dedicated resources to sensing improves estimation accuracy, but reduces communication spectral efficiency.
Accordingly, the communication--sensing tradeoff can be viewed as a multi-dimensional balance among communication capacity, resource occupancy, and sensing accuracy.

\subsection{Tradeoff between Sensing and RF Performance}
The sensing--RF tradeoff emerges because the waveform basis strongly influences both sensing characteristics and RF‑domain behavior.
For linearly modulated signals, the chosen orthogonal shapes the signal's energy distribution in the time and frequency domains.
Sensing benefits from waveforms with favorable correlation properties and low sidelobe levels, while RF power amplifiers favor well‑behaved time‑domain envelopes with restricted power fluctuations.
This creates an intrinsic conflict between the sensing performance and the efficiency of the power‐amplifier.

This tradeoff can be investigated through waveform-basis design by comparing the time‑ and frequency‑domain power distributions associated with different orthogonal bases and characterizing the resulting Pareto frontier.
OFDM typically provides satisfactory ranging performance but suffers from severe envelope fluctuations.
By contrast, single‑carrier waveforms achieve higher power‑amplifier efficiency at the expense of sensing capability.
Accordingly, the sensing--RF tradeoff is fundamentally tied to the intrinsic structure of the waveform basis, instead of mere parameter tuning.

\subsection{Tradeoff between Communication and RF Performance}
The communication--RF tradeoff stems from the conflict between retaining the inherent characteristics of communication signals and mitigating power fluctuations.
To guarantee reliable transmission and receiver compatibility, communication systems impose constraints on predefined modulation formats, spectral masks, and signal statistics.
In comparison, RF front‑ends prefer low‑PAPR signals for improved power‑amplifier efficiency.
Nevertheless, aggressive peak‑suppression operations distort the transmit signal, thus deteriorating the BER and EVM.

Existing low‑PAPR techniques can be broadly categorized into distortion‑based and distortion‑less approaches.
Clipping, filtering, and companding feature low implementation complexity, but introduce nonlinear distortion.
By contrast, distortion‑less schemes mitigate PAPR via intentional modification of transmitted signals, typically at the expense of additional side information overhead and increased computational complexity.
Accordingly, these techniques strike a balance among PAPR reduction capability, communication performance, and implementation complexity.
Alternatively, waveform‑basis designs exemplified by DFT‑s‑OFDM enhance envelope stability while incurring a relatively moderate performance penalty in communication.

\section{Design Methods of ISAC Signal}

\begin{figure*}[!h]
\centering
\includegraphics[width=0.82\textwidth]{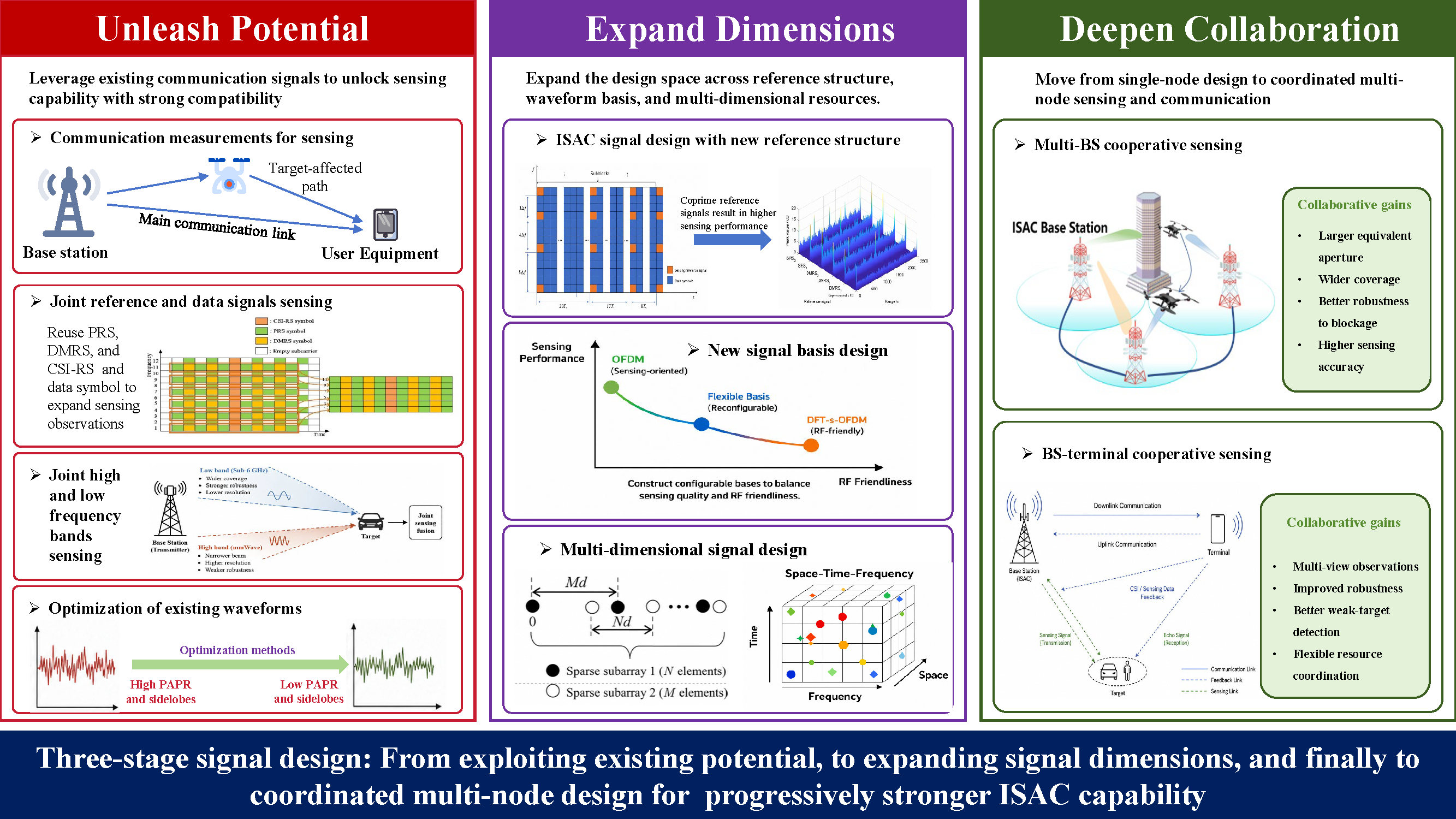}
\caption{Three-stage roadmap for ISAC signal design.}
\label{fig:fig2}
\end{figure*}

ISAC signal design evolves through three stages, as shown in Fig.~\ref{fig:fig2}.
Stage 1, termed ``Unleash Potential'', focuses on exploiting the capabilities of existing waveforms for ISAC.
Stage 2, termed ``Expand Dimensions'', expands the signal design space by developing new waveform bases.
Stage 3, termed ``Deepen Collaboration'', extends ISAC signal design from single-node operation to multi-node cooperation.

\subsection{Stage 1 (\textbf{Unleash Potential}): Leveraging Existing Waveforms}

Existing communication signals can be reused for sensing with only minor modifications to existing hardware and frame structures.
Representative approaches include utilizing communication measurements for sensing, jointly exploiting reference and data signals, jointly utilizing high- and low-frequency bands, and optimizing existing waveforms.

\subsubsection{Utilization of Communication Measurements for Sensing}

Communication measurements obtained during signal reception and processing can be used for sensing, as the presence and motion of targets induce measurable fluctuations in channel responses, received power, delay, Doppler shift, angle, and beam‑domain measurements.
Although these measurements are initially acquired for 
communication‐oriented tasks, including channel estimation, beam management, and mobility management, they can provide informative cues for target detection, localization, and tracking without requiring dedicated sensing signals.
Learning‑driven methods can further extract target-related variations from these measurements.
However, reliably separating target-related variations from these measurements remains a key challenge.

\subsubsection{Joint Utilization of Reference and Data Signals}

Existing 5G NR positioning reference signals (PRSs), demodulation reference signals (DMRSs), and channel state information reference signals (CSI-RSs) employ known sequences with desirable correlation properties, which make them suitable for sensing.
The joint utilization of multiple reference signals improves the continuity of sensing observations in the frequency and time domains, thereby enhancing the accuracy of range and velocity estimation~\cite{wei2024multiple}.
However, their sensing performance is constrained by predefined frame structures and sparse resource allocation. 
To address this issue, compressed sensing can recover sensing information from incomplete observations while balancing sensing accuracy and communication overhead. 
In addition, data signals occupy more time-frequency resources and can provide additional sensing observations. 
A practical approach is to obtain coarse channel and target estimates from reference signals and then refine them using data signals.
However, errors in data demodulation can propagate to sensing estimation, while iterative processing increases computational complexity.
Model-driven learning can alleviate this issue by unfolding iterative processing into a lightweight neural network.

\subsubsection{Joint Utilization of High- and Low-Frequency Bands}

Joint sensing across high- and low-frequency exploits their complementary propagation characteristics while reusing existing waveforms~\cite{Liu2025CA}.
Coordinated optimization of bandwidth, subcarrier spacing, sensing‑symbol allocation, and cyclic‑prefix length can improve sensing performance under a fixed resource budget.
However, this approach introduces considerable signaling overhead and latency in dynamic scenarios.
A lightweight library of predefined waveform configurations enables the selection of an appropriate waveform modes based on channel conditions and sensing requirements, thereby avoiding repeated online optimization.

\subsubsection{Optimization of Existing Waveforms}

Existing communication waveforms can be modified to enhance their sensing or RF performance. 
OFDM offers high spectral efficiency and favorable average ranging sidelobes even with random communication symbols.
However, its high PAPR reduces power‑amplifier efficiency, particularly in uplink ISAC~\cite{koivunen2024multicarrier}.
Phase optimization, companding, and statistical waveform shaping can mitigate PAPR while largely retaining correlation properties and communication performance.
By contrast, single-carrier waveforms produce a smoother signal envelope and provide higher RF efficiency, but generally exhibit higher ranging sidelobes.
Symbol deviations constrained by EVM can optimize the periodic auto‑correlation function (P‑ACF), thus reducing both integrated and peak sidelobe levels.
Iterative clipping and filtering serves as a low‑complexity approximate solution.
Thus, traditional waveform optimization offers a backward‑compatible approach, but must balance signal distortion, EVM budgets, and practical implementation constraints.

\subsection{Stage 2 (\textbf{Expand Dimensions}): Developing New Waveform Bases}

Beyond exploiting the existing communication waveforms, new ISAC signal designs reconfigure reference‑signal structures, waveform bases, and multi-dimensional resource allocation, thereby enabling more flexible tradeoffs among communication, sensing, and RF performance.

\subsubsection{New Reference-Signal Design}

New reference-signal designs modify pilot distributions and sequence structures. 
Coprime pilot patterns mitigate the range and velocity ambiguities caused by uniform sampling~\cite{Li2025Sparse}.
Combining Zadoff-Chu sequences with index modulation enables information to be encoded through pilot indices or cyclic shifts, eliminating the need for dedicated sensing signals.
These designs provide additional degrees of freedom to balance the accuracy of the estimation, the unambiguous range, and the spectral efficiency.
However, irregular pilot may reduce the effectiveness of conventional two-domensional (2D) FFT-based processing and increase complexity, thus motivating the development of low‑complexity sensing algorithms.

\subsubsection{New Waveform Base Design}\label{se3-2-2}

The waveform basis shapes the time‑frequency distribution of communication symbols, thus influencing both sensing capabilities and RF performance.
Although OFDM exhibits favorable ranging sidelobe, it exhibits severe envelope fluctuations and a high PAPR.
By contrast, single‑carrier waveforms, such as DFT‑s‑OFDM, typically reduce PAPR but may provide less favorable sensing performance~\cite{Long2026}.
Novel orthogonal or configurable bases enable tunable tradeoffs between ranging sidelobes and time‑domain power fluctuations, resolving the sensing‑RF conflict from the perspective of waveform structure.
However, deploying such waveform bases may require modifications to synchronization, equalization, channel estimation, and transceiver implementations.
Key research directions include characterizing the Pareto frontier, constructing low‑complexity waveform bases, and designing compatibility-oriented configurable waveforms.

\subsubsection{multi-dimensional Signal Design}

multi-dimensional signal design coordinates resource allocation across spatial, temporal, and frequency, thereby expanding the design space beyond conventional optimization confined to the time-frequency domain.
The coordinated design of array element activation, symbol placement, and subcarrier allocation can improve sensing resolution, unambiguous range, and multi‑target separability.
Nonuniform resource allocation can increase the effective sensing aperture without incurring proportional resource overhead.
However, coupled space‑time‑frequency variables introduce intricate constraints and place greater demands on system calibration, synchronization, and real‑time processing capabilities.
Promising future research directions include developing unified sampling models and sparse resource structures, as well as performing joint optimization under communication constraints.

\subsection{Stage 3 (\textbf{Deepen Collaboration}): Extending Signal Design to Multi-node Cooperation}

The first two stages focus mainly on single‑node signal design.
In contrast, cooperative sensing requires joint waveform design, resource allocation, and transceiver coordination between multiple nodes to exploit spatial diversity and extend the sensing coverage.

\subsubsection{Signal Design for Multi-BS Downlink Cooperation}
\label{multi-BS cooperation}

Multi-BS downlink cooperation can expand sensing coverage, increase the effective sensing aperture, and improve weak-target detection.
However, it also introduces severe inter-BS interference, complex signal coupling, and additional energy consumption.
An effective solution is to jointly optimize the operating modes and transceiver parameters of the BS. 
In this framework, distributed BSs are adaptively assigned to operate as transmitters or sensing receivers according to their respective sensing contributions.
BS--user association, transmit beamforming, and receive filtering are then jointly optimized subject to communication, sensing, and power constraints.

Separating echoes from multiple transmitting BSs is another critical challenge.
Orthogonal time-frequency resource allocation can prevent inter-BS signal mixing but inevitably reduces spectral efficiency.
By contrast, code-domain orthogonality techniques, such as space-time block coding (STBC), enable efficient resource reuse.
Specifically, a centralized baseband unit generates orthogonally coded OFDM signals for transmission by distributed BSs.
The sensing receivers then apply space-time decoding to separate superimposed echoes and fuse monostatic and bistatic sensing observations \cite{11359065}.
However, target-induced Doppler shifts may impair code orthogonality, while scaling to larger BS clusters requires longer code blocks, creating additional implementation challenges.

\subsubsection{Signal Design for BS--UE Cooperation}

BS--UE cooperation enables UE-assisted sensing, distributed echo acquisition, and extended sensing coverage.
Compared to multi-BS cooperation, signal design for BS--UE cooperation is subject to tighter constraints arising from the limited transmit power, power‑amplifier efficiency, and hardware complexity of UEs.
Therefore, low‐PAPR and low‑complexity signals, together with adaptive power control, are required.

Resource coordination presents another major challenge because distributed UEs differ in their communication requirements, sensing contributions, and channel conditions.
Dynamic UE selection and resource scheduling can jointly adapt waveform parameters, transmit power, and time‑frequency resource allocation based on UE status, sensing tasks, and target locations.

\section{Three-Stage Evolution of ISAC Signal Design Under Standardization}

\begin{figure*}[htbp]
\centering
\includegraphics[width=0.8\textwidth]{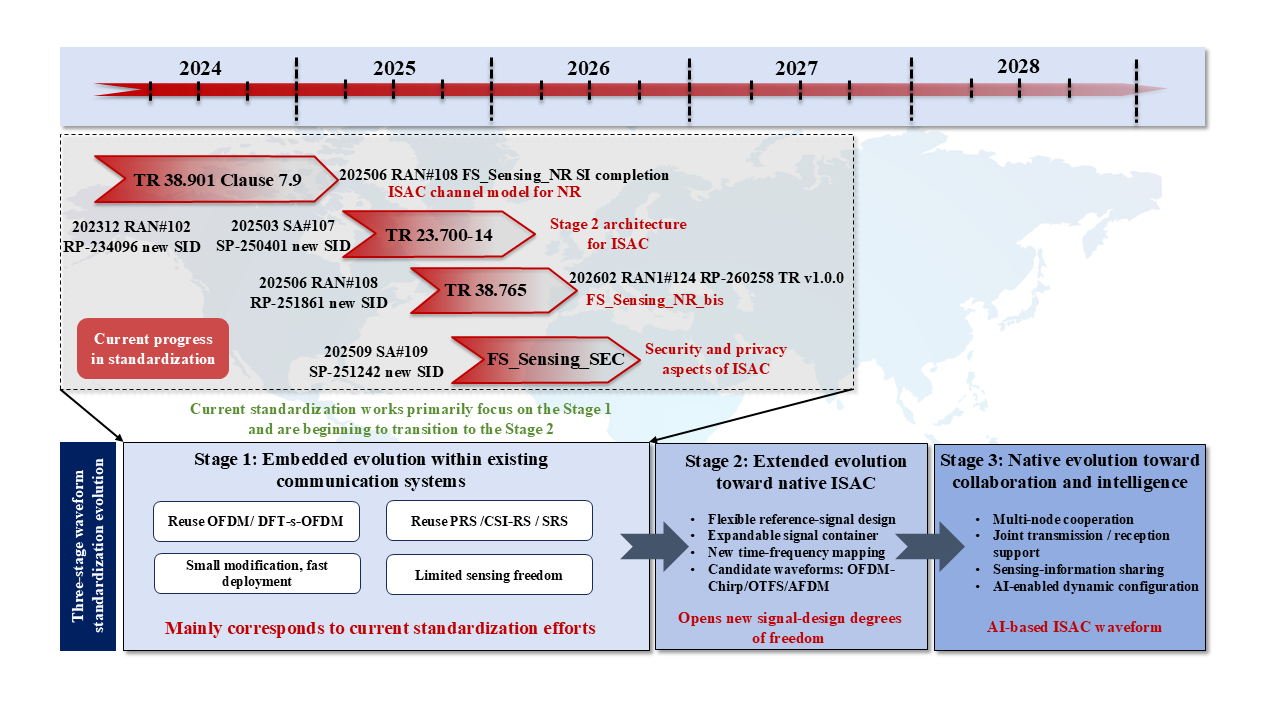}
\caption{Three-stage evolution of ISAC signal design under standardization.}
\label{fig:fig3}
\end{figure*}

Flexible ISAC signal design supports adaptation to diverse communication, sensing, and RF requirements, while standardization emphasizes uniformity and interoperability.
Accordingly, accommodating such flexibility within a unified standardization framework represents a key challenge for practical deployment.
To balance performance improvements with backward compatibility, ISAC signal standardization can adopt a three‑stage evolutionary roadmap: 
evolutionary embedding into existing communication systems, evolutionary extension toward native ISAC, and evolution for collaboration and intelligence, as shown in Fig.~\ref{fig:fig3}.

\subsection{Stage 1: Embedded Evolution within Existing Communication Systems}

The first stage introduces sensing functions while maintaining the fundamental physical‑layer architecture of existing cellular systems.
Existing waveforms, including OFDM and DFT‑s‑OFDM, as well as reference signals such as PRSs, CSI-RS and sounding reference signals (SRSs), can be reused for sensing.
Pilot enhancement, joint processing of reference and data signals, and flexible time‑frequency configuration enable basic sensing capabilities while maintaining compatibility with the existing communication architecture.

This approach aligns with the 3GPP Release 20 studies on ISAC within the NR framework~\cite{3GPP_TR_38_765}. 
Consequently, the key standardization tasks are to identify traditional sensing signals and to specify the associated configuration, measurement, and reporting mechanisms, rather than to define an entirely new waveform~\cite{3GPP_TS_23_137}.
This approach requires only minor system modifications, facilitating rapid deployment and field trials.

However, sensing performance remains constrained by the bandwidth, periodicity, resource allocation, and signal structures originally optimized for communication.
Consequently, this stage is suitable for basic sensing services, but cannot fully satisfy the requirements for high‑resolution, high‑mobility, or continuous sensing.
Therefore, a transition toward more native ISAC designs is necessary.

\subsection{Stage 2: Extended Evolution Toward Native ISAC}

In the second stage, signal design evolves from reproposing communication‑oriented signals for auxiliary sensing to developing a native ISAC signal.
Potential research directions include flexible reference-signal patterns, sensing-oriented time-frequency mappings, hybrid signal structures, and new waveform bases.
Representative examples include non-uniform reference signals and waveforms designed for high-mobility scenarios, such as OTFS and AFDM~\cite{rou2024orthogonal}.
Rather than standardizing a single fixed waveform, a more practical strategy is to define a configurable ISAC signal framework that supports diverse signal structures and parameter sets tailored to different services and deployment scenarios.

Existing 3GPP studies have investigated representative sensing scenarios, service requirements, and performance metrics~\cite{3GPP_TR_38_765,3GPP_TS_23_137}. 
These studies provide a foundation for incorporating greater flexibility into waveform and resource configuration in future standards.
However, introducing additional degrees of freedom into signal design poses new challenges in synchronization, channel estimation, equalization, RF implementation, and backward compatibility.
Therefore, the primary objective of this stage is not simply to replace existing waveforms but to extend the physical‑layer framework while ensuring manageable implementation complexity and compatibility with existing systems.

\subsection{Stage 3: Evolution toward Collaboration and Intelligence}

The third stage focuses on cooperative ISAC.
Standardization frameworks must therefore support cross-node synchronization, coordinated transceiving, sensing information fusion, joint resource management, and task-adaptive configuration.
Standards need to define the interfaces, constraints, and operating procedures required for cooperative signal design. 
As part of Release 20, 3GPP has initiated systematic studies of system-level ISAC,
including network architecture design, end-to-end sensing procedures, and the delivery of sensing results to vertical industry applications~\cite{3GPP_TR_23_700_14,3GPP_TR_23_700_15}.
The main challenge is to balance dynamic adaptation with interoperability. 
Overly rigid signal definitions and optimization rules restrict the flexible optimization capability of ISAC systems, while insufficient normative constraints may impede multi-node cooperative operations.
In the future, standards should therefore define configurable parameter ranges, mandatory performance and signaling constraints, and standardized interaction procedures for cooperative sensing.

\section{Simulation Experiments}

This section first evaluates the waveform-basis optimization method introduced in Section~\ref{se3-2-2} for balancing sensing and RF performance.
Then, it investigates the STBC-based multi-BS cooperative sensing method presented in Section~\ref{multi-BS cooperation}.

\subsection{Waveform-Base Optimization}

\begin{figure}[htbp]
\centering
\includegraphics[width=0.7\columnwidth]{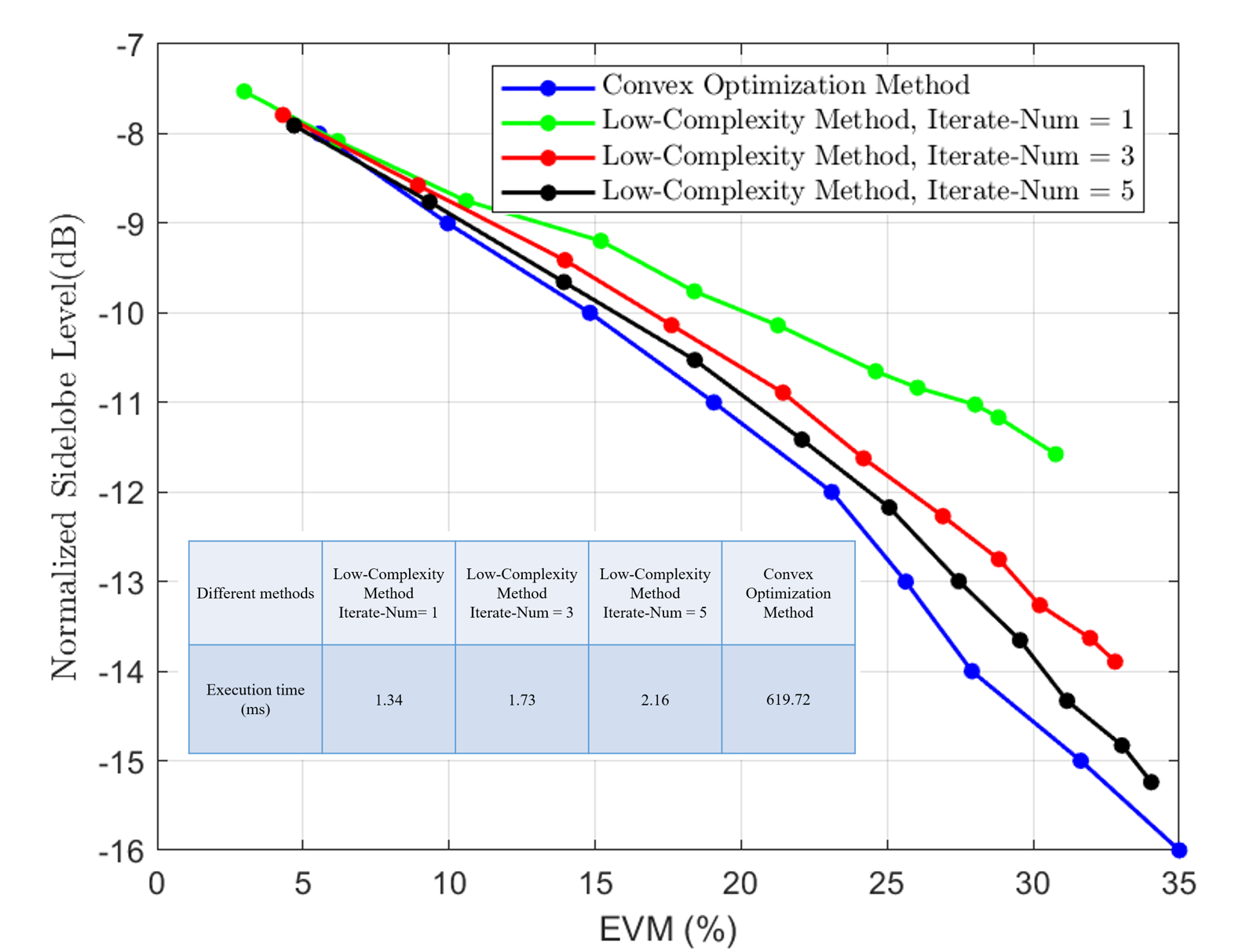}
\caption{EVM-SL tradeoff of the optimization-based and low-complexity methods.}
\label{fig:EVM_PSL_tradeoff}
\end{figure}

The number of subcarriers and the oversampling factor are set to 128 and 4, respectively. 
Fig.~\ref{fig:EVM_PSL_tradeoff} compares the level of the EVM-sidelobe achieved by the convex optimization approach and the low‑complexity P‑ACF clipping‑and‑filtering method.
The simulation results reveal that the low‑complexity method can closely approach the performance of convex optimization with merely $5$ iterations.
Execution-time comparisons further demonstrate that it achieves comparable sidelobe suppression at a substantially low computational cost. 
These results indicate that efficient waveform optimization can improve sensing‑oriented sidelobe suppression while maintaining communication distortion within acceptable limits.

\subsection{Multi-BS Cooperative Sensing}

\begin{figure}[htbp]
\centering
\includegraphics[width=0.9\columnwidth]{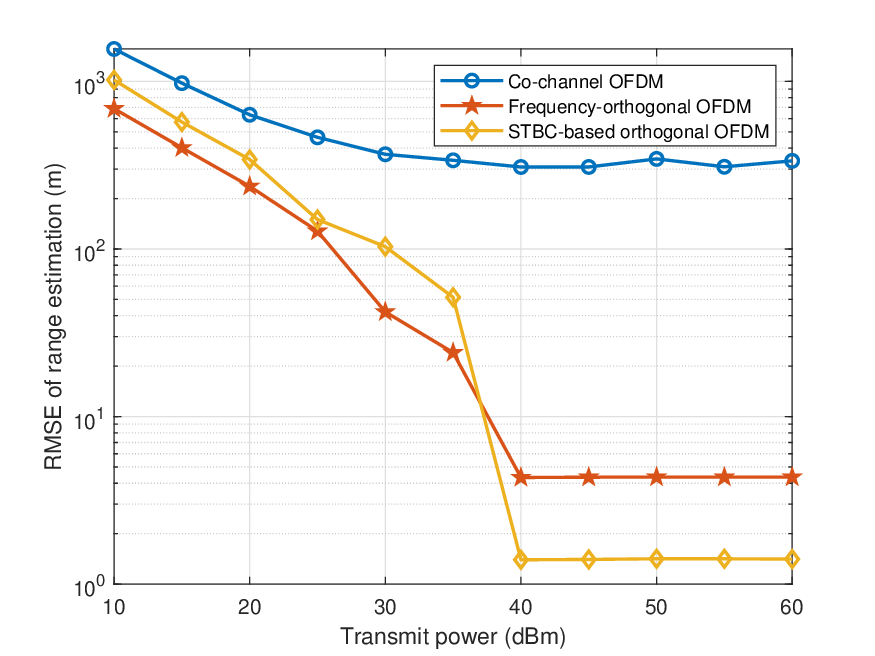}
\caption{Range-estimation RMSE under different multi-BS cooperative waveforms.}
\label{fig:cooperative_ranging_rmse}
\end{figure}

The simulation considers three BSs with an inter-site distance of $300$ m and a height of $30$ m. 
The carrier frequency, subcarrier spacing, and number of OFDM subcarriers are set to $2.6$ GHz, $30$ kHz, and 1024, respectively. 
Four consecutive OFDM symbols are transmitted every $2.5$ ms for sensing. 
The target radar cross‑section (RCS) is set to $0.01$ m$^2$, and target locations are randomly generated in the cell-center region over 10,000 Monte Carlo trials.

Fig.~\ref{fig:cooperative_ranging_rmse} compares three cooperative signal design schemes. 
In \emph{co-channel OFDM}, all BSs reuse the same time‑frequency resources, where echoes originating from adjacent BSs serve as interference.
In \emph{frequency-orthogonal OFDM}, different BSs are allocated disjoint frequency resources to eliminate inter‑BS interference.
In the proposed \emph{STBC-based orthogonal OFDM}, STBC enables different BSs to reuse the same time-frequency resources, while their echoes are separated through space-time block decoding \cite{11359065}.
The simulation results reveal that inter-BS interference significantly degrades ranging performance. 
Compared with \emph{frequency-orthogonal OFDM}, the proposed \emph{STBC-based orthogonal OFDM} achieves superior ranging accuracy by suppressing inter-BS interference while fully exploiting the available bandwidth. 
Under a sensing bandwidth of $30$ MHz, it achieves a range-estimation RMSE of approximately $1.4$ m.

\section{Conclusion}

This article investigates ISAC signal design for 6G from the perspectives of performance metrics, signal design, and standardization. 
First, a multi-dimensional performance metric framework is established to jointly characterize communication, sensing, and RF performance.
Building on this framework, a three‑stage ISAC signal design paradigm comprising \textsc{Unleash Potential}, \textsc{Expand Dimensions}, and \textsc{Deepen Collaboration} is presented to trace the evolution from exploiting traditional communication signals, through developing novel waveform bases and multi-dimensional signal structures, to enabling multi-node cooperation.
Consequently, a standardization roadmap that evolves from backward compatibility with existing communication systems toward native ISAC is elaborated.
These analyzes provide a systematic foundation for future ISAC signal design and standards development in 6G networks.

\bibliographystyle{IEEEtran}
\bibliography{reference}
\end{document}